# Precise measurement of $\theta_{13}$ at Daya Bay

M. –C. Chu

*Department of Physics, The Chinese University of Hong Kong, Shatin, N.T., Hong Kong.*
*On behalf of the Daya Bay Collaboration*

The Daya Bay Reactor Neutrino Experiment is designed to determine the yet unknown neutrino mixing angle $\theta_{13}$ by measuring the disappearance of electron antineutrinos from several nuclear reactor cores, using multiple underground detectors at different baselines to minimize systematic errors and to suppress the cosmogenic background. The civil construction has begun since October 2007, enabling first commissioning data in 2009, and full data taking will begin in late 2010. The planned sensitivity in $\sin^2(2\theta_{13})$ of better than 0.01 at 90% CL will be achieved in three years of data-taking. I will present an overview and current status of the experiment.

The Standard Model of Particle Physics (SM) had assumed that all three types of neutrinos were massless. The recent observation of neutrino oscillation, however, has unequivocally demonstrated that at least two types of neutrinos are massive and can mix, thereby showing that the SM is incomplete [1]. The smallness of the neutrino masses ($< 2$ eV) and the two large mixing angles measured have thus far provided important clues and constraints to physics beyond the SM. The third, yet undetermined, mixing angle, $\theta_{13}$, is surprisingly small; the current experimental bound is $\sin^2 2\theta_{13} < 0.17$ at 90% confidence level for the nominal value of the mass-squared difference $\Delta m^2_{31} = 2.5 \times 10^{-3} \text{eV}^2$ [2]. Besides completing the picture of neutrino mixing [3], it is important to measure this angle to provide further insight on how to extend the SM. The value of $\theta_{13}$ is also crucial for searching for CP violation in the lepton sector [4], described by the parameter $\delta_{CP}$, which might explain why the present Universe is matter-dominated. The matter effect can be used to determine the mass hierarchy (order of the mass eigenstates), but it also depends on the value of $\theta_{13}$ [5]. Exploiting the low-energy electron antineutrinos, a reactor-based experiment can measure $\theta_{13}$ without any ambiguity, and is complementary to the long-baseline, accelerator-based experiments that suffer from the matter effect and potential CP violation when the value of $\theta_{13}$ is extracted from the measurements [6]. The goal of the Daya Bay experiment is to reach a sensitivity of 0.01 or better in $\sin^2 2\theta_{13}$ at 90% confidence level.

Nuclear reactors are abundant and stable sources of electron antineutrinos [7]. Reactor neutrino spectra and flux have been determined to an accuracy of 2% [6]. Reactor-based neutrino experiments are disappearance experiments because the $\overline{\nu}_{\mu}$ coming from oscillation does not have enough energy to produce a $\mu$ through the charged-current process. The survival probability for reactor electron antineutrinos (shown in Fig. 1) depends only on $\theta_{13}$ and $\Delta m^2_{31}$ for baseline up to a couple kilometers. $\sin^2 2\theta_{13}$ is given exactly by the amplitude of the small oscillations. Therefore, the optimal baseline to measure the value of $\theta_{13}$ is at the first minimum in the survival probability near 2 km. The systematic uncertainties can be greatly suppressed or totally eliminated in a *relative* measurement between two identical detectors having the same efficiency positioned at two different baselines [8]. The near detector close to the reactor core is used to establish the flux and energy spectrum of the antineutrinos. The value of $\sin^2 2\theta_{13}$ can be measured by comparing the antineutrino flux and energy distribution observed with the far detector to those of the near detector after scaling with the inverse-square law.

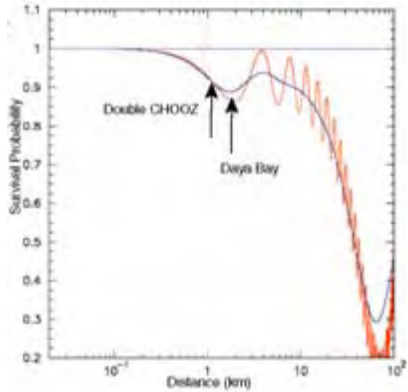

Fig. 1. Survival probability of electron antineutrinos vs. distance from their source. The arrows refer to the locations of the far detector in the Double Chooz and Daya Bay experiments. The amplitude of the small oscillations in the red curve is taken at the Chooz limit assuming $\Delta m^2_{31}= 2.5 \times 10^{-3}$ eV$^2$. The red curve is for 4 MeV antineutrinos, and the blue curve is the result smeared by the expected reactor antineutrino energy spectrum.

The Day Bay nuclear power complex will be among the five most powerful reactor complexes in the world and generate 17.4 GW$_{th}$ of power by early 2011, when a third pair of reactor cores is put into operation, joining the two pairs of existing reactor cores. The site is located adjacent to mountainous terrain, ideal for constructing underground detector laboratories that are well shielded from cosmogenic backgrounds. The basic experimental layout of Daya Bay consists of three underground experimental halls, one far and two near, linked by horizontal tunnels. Figure 2 shows the detector module deployment at these sites [9].

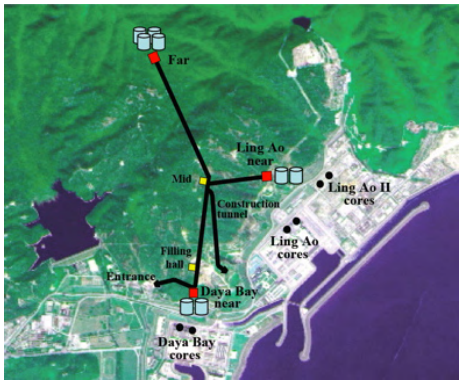

Fig. 2. Default configuration of the Daya Bay experiment, optimized for the best sensitivity in $\sin^2 2\theta_{13}$. Four detector modules are deployed at the far site and two each at each of the near sites.

Antineutrinos are detected by an organic liquid scintillator (LS) via the inverse beta-decay reaction. The prompt positron signal and delayed neutron-capture signal are combined to define a neutrino event with timing and energy requirements on both signals. When Gadolinium (Gd) is loaded into the LS, neutrons are captured with a huge cross section and short capture time (~30 $\mu$s), releasing 8 MeV of $\gamma$-ray energy. The signals can thus be cleanly separated from the background from accidental coincidences. Each three-zone detector module (Fig. 3) will have 20 metric tons of 0.1% Gd-doped LS in the inner-most, antineutrino target zone. A second zone, separated from the target and outer buffer zones by transparent acrylic vessels, will be filled with undoped LS for capturing gamma rays that escape from the target, thereby improving the antineutrino detection efficiency. A total of 192 PMT's are arranged along the circumference of the stainless steel tank in the outer-most zone, which contains mineral oil to attenuate gamma rays from trace radioactivity in the PMT glass and nearby materials. The AD modules are submerged in a water pool that shields the modules from ambient radiation and spallation neutrons (Fig. 3). The water pool is segmented into two partitions using reflective Tyvek sheets. PMTs are installed in the partitions to collect Cherenkov photons produced by cosmic muons in the water. Above the pool the muons are tracked with a detector made of light-weight resistive-plate chambers (RPCs), which offer good performance and excellent position resolution at low cost.

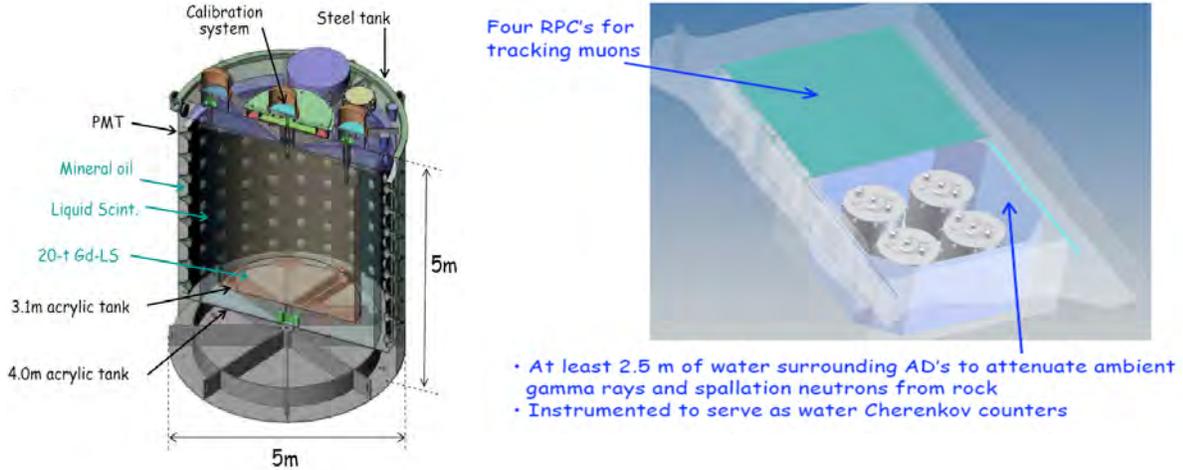

Fig. 3. (Left) Cross sectional slice of a 3-zone AD module showing the acrylic vessels holding the Gd-LS at the center (20 T), LS between the acrylic vessels (20 T) and mineral oil (40 T) in the outer region. The PMTs are mounted on the inside walls of the stainless steel tank. (Right) AD modules are shielded by a 1.5 m-thick active water Cherenkov buffer. Surrounding this buffer is another 1-meter of water Cherenkov tanks serving as muon trackers. The muon system is completed with RPCs at the top.

Table 1 summarizes the signals and background for each of the three experimental halls of the Daya Bay Reactor Neutrino Experiment [9]. With three years of data, the sensitivity contours of the Daya Bay experiment in the $\sin^2 2\theta_{13}$ versus $\Delta m^2_{31}$ plane are shown in Fig. 4. The green shaded area shows the 90% confidence region of $\Delta m^2_{31}$ determined by atmospheric neutrino experiments. Assuming four 20-ton modules at the far site and two 20-ton modules at each near site, the statistical uncertainty is around 0.2%. The sensitivity of the Daya Bay experiment with this design can achieve the challenging goal of 0.01 with 90% confidence level over the entire allowed (90% CL) range of $\Delta m^2_{31}$. At the best fit $\Delta m^2_{31}$= 2.5×10$^{-3}$ eV$^2$, the sensitivity is around 0.008 with 3 years of data.

|  | Daya Bay Near | Ling Ao Near | Far Hall |
|---|---|---|---|
| Baseline (m) | 363 | 481 from Ling Ao<br>526 from Ling Ao II | 1985 from Daya Bay<br>1615 from Ling Ao |
| Overburden (m) | 98 | 112 | 350 |
| Radioactivity (Hz) | <50 | <50 | <50 |
| Muon rate (Hz) | 36 | 22 | 1.2 |
| Antineutrino Signal (events/day) | 840 | 740 | 90 |
| Accidental Background/Signal (%) | <0.2 | <0.2 | <0.1 |
| Fast neutron Background/Signal (%) | 0.1 | 0.1 | 0.1 |
| $^8$He+$^9$Li Background/Signal (%) | 0.3 | 0.2 | 0.2 |

Table 1 Summary of signals and background at the various experimental halls.

Insert PSN Here

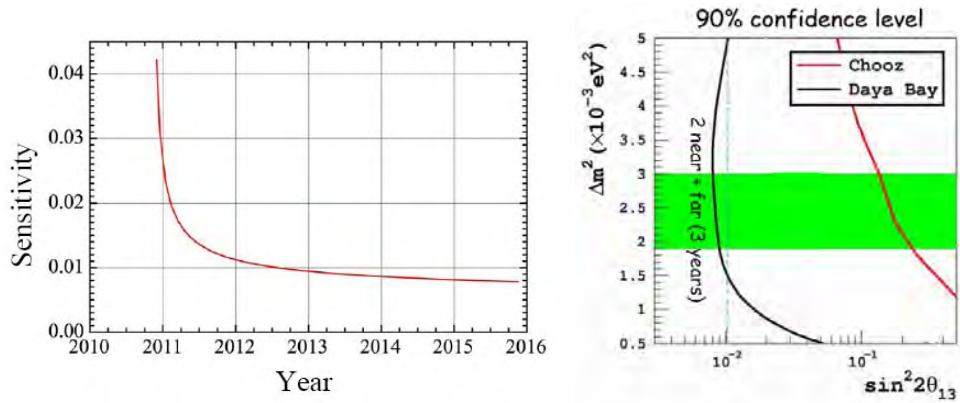

Fig. 4 Expected sensitivity to $\sin^2 2\theta_{13}$ at Daya Bay with 3 years of data taking. (Right) Contours at 90% C.L. with 3 years of data taking of the Daya Bay experiment in the $\sin^2 2\theta_{13}$ versus $\Delta m^2_{31}$ plane. The red line shows the current limit.

Civil construction at the Daya Bay experimental site has already begun since the groundbreaking on October 13, 2007. Almost all subsystem prototypes, including those of the anti-neutrino detector, muon system, calibration system, and front-end and readout electronics have been completed. All long-lead orders have been initiated. The Daya Bay Reactor Neutrino Experiment is moving full-steam ahead to reach a sensitivity of 0.01 or better in $\sin^2 2\theta_{13}$.

## Acknowledgments

This work was supported in part by the the Chinese Academy of Sciences, the National Natural Science Foundation of China (Project numbers 10225524, 10475086, 10535050 and 10575056), the Ministry of Science and Technology of China, the Guangdong provincial goverment, the Shenzhen Municipal government, the China Guangdong Nuclear Power Group, the Research Grants Council of the Hong Kong Special Administrative Region of China (Project numbers 2300017, 400805 and 400606), the Chinese University of Hong Kong Focused Investment Grant (Project number 3110031), the United States Department of Energy (Contracts DE-AC02-98CH10886, DE-AS02-98CH1-886, DE-FG02-92ER40709 and DE-FG02-91ER40671 and Grant DE-FG02-88ER40397), the U.S. National Science Foundation (Grants PHY-0653013, PHY-0650979, PHY-0555674 and NSF03-54951), the University of Houston (GEAR Grant number 38991), the University of Wisconsin and the Ministry of Education, Youth and Sports of the Czech Republic (Project numbers MSM0021620859 and LC527).

## References

[1] C. Amsler et al. (Particle Data Group), Physics Letters **B667**, 1 (2008).

[2] G.L. Fogli, E. Lisi, A. Marrone, and A. Palazzo, and A.M. Rotunno, [arXiv:hep-ph/0806.2649].

[3] Z. Maki, M. Nakagawa, and S. Sakata, Prog. Theor. Phys. 28, 870 (1962); B. Pontecorvo, Sov. Phys. JETP 26, 984 (1968); V.N. Gribov and B. Pontecorvo, Phys. Lett. B28, 493 (1969).

[4] V. D. Barger, D. Marfatia and K. Whisnant, "Neutrino Superbeam Scenarios At The Peak," in Proc. of the APS/DPF/DPB Summer Study on the Future of Particle Physics (Snowmass 2001) ed. N. Graf, eConf C010630, E102 (2001) [arXiv:hep-ph/0108090].

[5] V. Barger, D. Marfatia and K. Whisnant, Phys. Lett. B560, 75 (2003) [arXiv:hep-ph/0210428]; P. Huber, M. Lindner and W. Winter, Nucl. Phys. B654, 3 (2003) [arXiv:hep-ph/0211300]; H. Minakata, H. Nunokawa and S. Parke, Phys. Rev. D68, 013010 (2003) [arXiv:hep-ph/0301210].

[6] C. Bemporad, G. Gratta, and P. Vogel, Rev. Mod. Phys. 74, 297 (2002); M. Diwan et al., hep-ex/0608023.

[7] P. Vogel and J. Engel, Phys. Rev. D39, 3378 (1989).

[8] L.A. Mikaelyan and V.V. Sinev, Phys.Atomic Nucl. 63 1002 (2000) [arXiv:hep-ex/9908047].

[9] For a detailed discussion of the Daya Bay experiment, see http://arxiv.org/abs/hep-ex/0701029v1.